\documentclass[12pt,a4paper,twoside]{article}
\usepackage{epsfig}

\newcommand{\be}{\begin{equation}}
\newcommand{\ee}{\end{equation}}
\newcommand{\beqs}{\begin{eqnarray}}
\newcommand{\eeqs}{\end{eqnarray}}

\expandafter\ifx\csname mathbbm\endcsname\relax

\else

\fi
\begin{document}
\begin{titlepage}
\begin{flushleft}  
       \hfill                      {\tt hep-th/9804141}\\
       \hfill                      UUITP-2/98\\
       \hfill                       April 1998\\
\end{flushleft}
\vspace*{3mm}
\begin{center}
{\LARGE Quarks, monopoles and dyons at large N
\\}
\vspace*{12mm}
{\large Ulf H. Danielsson\footnote{E-mail: ulf@teorfys.uu.se}$^{,a}$ \\
\vspace*{5mm}
\large Alexios P. Polychronakos\footnote{E-mail:
poly@teorfys.uu.se}$^{,a,b}$ \\
\vspace*{5mm}
{\em (a) Institutionen f\"{o}r teoretisk fysik \\
Box 803 \\
S-751 08  Uppsala, Sweden \/}\\
\vspace*{5mm}
{\em (b) Physics Department, University of Ioannina \\
45110 Ioannina, Greece\/}\\}
\vspace*{15mm}
\end{center}

\begin{abstract}
We study a system of external particles of various charges in
${\cal N} =4$ 
super Yang-Mills in the large $N$ limit at finite temperature. 
We demonstrate that at high enough temperature partial or
complete screening of the particles can occur. At zero temperature
the total electric or magnetic charge cannot be screened, while
higher multipole moments of these charges can be screened.
The specific case of a quark, a monopole and a dyon is worked out
and the above properties are verified.
We also discuss the free energy of isolated particles and show that
their entropy is independent of the temperature.

\end{abstract}

\end{titlepage}

\section{Introduction}

The recently proposed connection between anti-de Sitter (AdS) supergravity
and conformal field theory at
its boundary \cite{Ma1,GKP,Wi1} (see also \cite{Pol}) permitted the calculation in the large-N
${\cal N}=4$ super Yang-Mills theories of
quantities that were previously inaccessible. In \cite{Ma2,RY} 
the temporal Wilson loop for a quark anti-quark pair at zero
temperature was calculated and their interaction energy thus deduced. 
This was later extended to finite temperature
in \cite{BISY,RTY} and to a system of a quark and a monopole at zero
temperature in
\cite{Mi}. Furthermore, a demonstration of the existence of a mass gap
in this theory was given in \cite{Wi2}. See \cite{theclub} for 
other recent results.

What makes the physics nontrivial is that for any nonzero temperature
the system is actually in a `high-temperature' phase \cite{Wi2}. 
This phase persists at zero temperature as one of two possible phases,
and is the one examined here and in the literature.
(The `low-temperature' confining phase would be obtained by taking the 
temperature to zero before taking the volume to infinity and is, 
arguably, physically less interesting \cite{Wi2}.)
As a result, isolated
quarks can exist even at zero temperature, signaling the
breakdown of $Z_N$ symmetry and the presence of a conformally
invariant screening in the gluon vacuum. 

In this paper we will investigate the general properties of a system 
of external particles in this theory, concentrating on issues of
screening and clustering. 
As a representative example we will study and explicitly work out the 
properties of a system consisting of three particles: a
quark, a monopole and a dyon at nonzero temperature. 
The charges are chosen such that the particles can form a singlet 
bound state, so that questions of clustering
versus screening as the system is heated up can be addressed. 
At sufficiently high temperature, 
one of the particles is screened from the 
other two and at even higher temperature we get complete screening of
all three. 
A similar analysis holds for more general
collections of particles and the features that generalize 
are pointed out.

The entropy of isolated particles is also calculated and shown to be 
independent of the temperature. This determines the effective internal 
states of external particles in this theory and raises the interesting 
issue of a microscopic derivation of these states. 

\section{The setup}

Consider the specific case of a quark, a monopole and a dyon.
For simplicity we will assume that the three particles are situated along
a line. As an example, we choose to put the dyon in the middle.

The euclidean metric of the non-extremal D3-brane is given by
\be
ds ^2 = \alpha ' \left(\frac{U^2}{R^2} \left\{(1-U_T^4/U^4)d\tau^2 +dx_i^2
\right\} +
R^2(1-U_T^4/U^4)^{-1} \frac{dU^2}{U^2}\right)
\ee
The horizon is at $U_T$ and has become the origin of
the euclidean coordinates.
Space is restricted to the $U \ge U_T$ domain.
The periodicity of euclidean time $\tau$, ensuring the absence of
conical singularities at $U=U_T$, is $\beta = 1/T = \pi R^2 / U_T$,
while
the coupling constant $g_{YM}$ of the corresponding large-$N$ theory 
is given as $g_{YM}^2 N = R^4$, (we have $g = g_{YM}^2 /4\pi$). 
We note that $R$ is dimensionless while
$U_T$ has dimensions of energy.

In the large-$N$, large-$g_{YM}^2 N$ domain the expectation value of the
Wilson-'t Hooft loops corresponding to the insertion of external
particles can be approximated by the classical minimum of the action
of the corresponding AdS string configuration.
Figure 1 shows the configuration that we have in mind for the three
particles.
Each is represented by a string, a $(1,0)$ string with tension $1/2\pi
\alpha '$ for
the quark, a $(0,1)$ string with tension $1/2\pi \alpha 'g$ for the
monopole
and a $(-1,-1)$ string with tension $\sqrt{1+1/g^2}/2\pi \alpha '$ for
the dyon. 
They connect
in a vertex at $U=U_0$. This kind of three string junctions has been
extensively 
discussed in the literature, \cite{3string}.

\begin{figure}  
\begin{center}                                                         
\epsfig{file=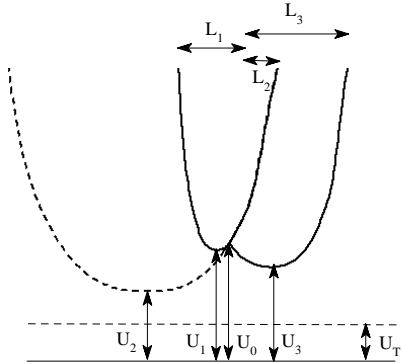,width=7 cm}                                               
     
\caption{ A configuration connecting a quark, a dyon and a
monopole.}
\end{center}
\end{figure}

The general shape of the strings is determined by locally minimizing
the world-sheet surface area, i.e.,
\be
S= \frac{T}{2\pi} \int dx \sqrt{(\partial _x U)^2 +(U^4 -U^4_T )/R^4} .
\ee
It is found that
\be
\frac{U^4-U_T^4}{\sqrt{(\partial _x U)^2 +(U^4-U_T^4)/R^4}} = R^2
\sqrt{U_i^4-U_T^4},
\ee
where $U_i$ are the minima of $U$ for the three strings respectively.
(In case of the dyon string it is the minimum of the extension of the
string past the vertex.) 

At the vertex, the forces exerted
by the three strings due to the tension must sum to zero. By using the
above expressions for the shapes of the strings we find 
the following conditions.
In the horizontal direction we find 
\be
y_1  =  \sqrt{1+1/g^2} y_2 + \frac{1}{g} y_3  , 
\label{condqdm1}
\ee
while in the vertical direction we find
\be
\sqrt{1-y_1^2} + \frac{1}{g} \sqrt{1-y_3^2} = \sqrt{1+1/g^2}
\sqrt{1-y_2^2} .
\label{condqdm2}
\ee
For convenience we have defined
\be
y_i^2 = \frac{U_i^4-U_T^4}{U_0^4-U_T^4} ,
\ee
It will also be convenient to define
\be
x_i = \frac{U_i^2}{U_0^2} = \sqrt{y_i^2 +\frac{U_T^4}{U_0^4}(1-y_i^2)}.
\ee
The conditions (\ref{condqdm1}) and (\ref{condqdm2})  
are solved by
\be
y_2=\frac{g y_1-\sqrt{1-y_1^2}}{\sqrt{1+g^2}}
\ee
and
\be
y_3=\sqrt{1-y_1^2} .
\ee
We will use $y_1$ as a parameter. Together with $U_0$ it parametrizes
all 
the configurations of the system (only the two relative distances
between the particles are relevant).

The free energy $F$ of the configuration is obtained by dividing the
total
worldsheet action of the above configuration by the inverse temperature.. 
This is the relevant quantity when considering
equilibrium conditions and transitions from meta\-stable states. The
mean energy $E$ of the configuration (in excess of the thermal vacuum),
on the other hand, is found by differentiating the total action with
respect to the inverse temperature, and would be relevant to local
nonequilibrium processes. We will mostly consider the free energy
in this work.

The worldsheet area is infinite, due to the branches extending to
$U=\infty$, and needs to be regularized. The origin of this is the infinite
mass of the heavy particles inserted \cite{Ma2}, which should be
subtracted. This corresponds to the free energy of three
isolated particles at zero temperature, calculated by evaluating the
worldsheet action of three free strings reaching down from infinity
to $U = 0$ in the extremal geometry ($U_T = 0)$. We can perform
this subtraction by
introducing a cutoff $\Lambda$ for $U$ in both the extremal and
the near-extremal case, keeping $R$ the same and introducing
the same periodicity $\beta$ to ensure the same asymptotic geometry
at $U = \Lambda$.
(A subtlety involving a slightly different periodicity in the two
cases, which appears when calculating the total world volume action
\cite{HP,Wi2}, is actually irrelevant here.) This should be contrasted
to the procedure used previously in \cite{BISY,RTY} where the subtraction 
was with a free string down to $U=U_T$ in the near-extremal space,
reducing the free energy of all isolated particles to zero 
at any temperature. We obtain for the
free energy, mean energy and entropy of an isolated particle
\be
F = - Q \sqrt{\pi g N} \, T ~,~~ E = 0 ~,~~ 
S = Q \sqrt{\pi g N}
\ee
where $Q=\sqrt{p^2 + q^2 /g^2}$ is the BPS mass of the particle
in quark mass units.
The result $E=0$ signifies that thermal vacuum polarization effects 
lead to no additional accumulation of energy around the particle.
Interestingly, we obtain a constant value for the entropy of a
particle, a result also valid at zero temperature. The above would imply 
that the presence of the particle, apart from a shift of the energy by
by a constant (which applies towards renormalizing the mass
of the particle), introduces a constant degeneracy factor $e^S$
corresponding, presumably, to the effective internal states of the 
particle in this phase.

With the above regularization, the free energy of our configuration
is found by adding a piece for each of the string segments, 
subtracting the contribution for three free strings reaching down 
from infinity to the horizon and adding the free energy of each
isolated particle. The result is:  
$$ 
F_{QDM}=\frac{U_1}{2\pi}  \int _1^{\infty}
\bigg(\frac{\sqrt{z^4-\frac{U_T^4}{U_1^4}}}{\sqrt{z^4-1}} -1\bigg)
-\frac{U_1}{2\pi}
+\frac{U_1}{2\pi}  \int _1^{1/\sqrt{x_1}}
\frac{\sqrt{z^4-\frac{U_T^4}{U_1^4}}}{\sqrt{z^4-1}}
$$ 
$$ 
+\frac{1}{g}\left(\frac{U_3}{2\pi} \int _1^{\infty}
\bigg(\frac{\sqrt{z^4-\frac{U_T^4}{U_3^4}}}{\sqrt{z^4-1}} -1\bigg) -
\frac{U_3}{2\pi}
+\frac{U_3}{2\pi}  \int _1^{1/\sqrt{x_3}}
\frac{\sqrt{z^4-\frac{U_T^4}{U_3^4}}}{\sqrt{z^4-1}}\right)
$$ 
\be 
+\sqrt{1+1/g^2}\left(\frac{U_2}{2\pi} 
\int _{1/\sqrt{x_2}}^{\infty}
\bigg(\frac{\sqrt{z^4-\frac{U_T^4}{U_2^4}}}{\sqrt{z^4-1}} -1\bigg)
-\frac{U_0}{2\pi}\right)   .
\ee

We also need the distances between the particles as indicated in the
figure. 
We find
\be
L_1 = \frac{R^2}{U_1} \sqrt{1-\frac{U_T^4}{U_1^4}} 
\bigg( \int _1^{\infty}
\frac{1}{\sqrt{z^4-\frac{U_T^4}{U_1^4}}\sqrt{z^4-1}}
+\int _1^{1/\sqrt{x_1}}
\frac{1}{\sqrt{z^4-\frac{U_T^4}{U_1^4}}\sqrt{z^4-1}}
\bigg)
\ee
\be
L_2 = sign(y_2 )\frac{R^2}{U_2} \sqrt{1-\frac{U_T^4}{U_2^4}} 
\int _{1/\sqrt{x_2}}^{\infty}
\frac{1}{\sqrt{z^4-\frac{U_T^4}{U_2^4}}\sqrt{z^4-1}}
\ee
and
\be
L_3 = \frac{R^2}{U_3} \sqrt{1-\frac{U_T^4}{U_3^4}} 
\bigg( \int _1^{\infty}
\frac{1}{\sqrt{z^4-\frac{U_T^4}{U_3^4}}\sqrt{z^4-1}}
+\int _1^{1/\sqrt{x_3}} \frac{1}{\sqrt{z^4-\frac{U_T^4}{U_3^4}}
\sqrt{z^4-1}}\bigg)  .
\ee
Note that we allow for the dyon to be to the right or to the left of the
vertex 
by being careful with 
the sign of $y_2$ or equivalently $L_2$. In the above expressions $U_i$ 
should be expressed as
a function of $y_1$ using the vertex conditions.

The above configuration will compete with three other possibilities
depicted in figure 2.
The first connects the quark and the monopole as in \cite{Mi}, with a
$(1,1)$ string reaching down from $U_0$ to the horizon. 
The vertex condition is
\be
y_1 = \frac{1}{g} y_3
\ee
and
\be
\sqrt{1-y_1^2} + \frac{1}{g} \sqrt{1-y_3^2} = \sqrt{1+1/g^2} . 
\ee
These conditions are solved by
\be
y_1=\frac{1}{\sqrt{1+g^2}}
\ee
\be
y_3=\frac{g}{\sqrt{1+g^2}} .
\ee
The second connects the quark and the dyon giving:
\be
y_1=\sqrt{1+1/g^2} y_2 
\ee
and
\be
\sqrt{1-y_1^2} + \frac{1}{g} = \sqrt{1+1/g^2} \sqrt{1-y_2^2} 
\ee
These conditions are solved by
\be
y_1 =1
\ee
\be
y_2=\frac{g}{\sqrt{1+g^2}}
\ee
Note that the quark string is always at its minimum at $U_0$.
(This is due to the fact that the $(1,0)$ string and the vertical
$(0,-1)$ string must join at right angles.)
Finally we have the dyon and the monopole with
\be
\sqrt{1+1/g^2} y_2 = \frac{1}{g} y_3
\ee
and
\be
\sqrt{1+1/g^2} \sqrt{1-y_2^2} =\frac{1}{g} \sqrt{1-y_3^2} +1 . 
\ee
These conditions are solved by
\be
y_2=\frac{1}{\sqrt{1+g^2}}
\ee
\be
y_3=1
\ee
Now it is the monopole string that always is at its minimum at $U_0$.

\begin{figure}  
\begin{center}                                                         
\epsfig{file=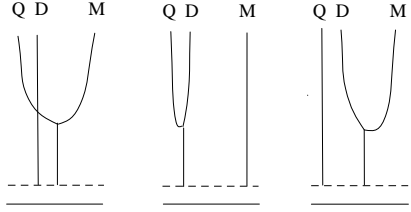,width=7 cm}                                               
     
\caption{ Three competing configurations.}
\end{center}
\end{figure}

We need the free energies for these configurations as well. 
At zero temperature these can be obtained as
functions of the distances between the particles. The strategy is then
to pick a value for $y_1$,
calculate the free energy for the QDM configuration, read 
off the corresponding distances, plug them into
the expressions for the paired configurations and compare. 
At non zero temperature we can not do that.
Instead we have to tune $U_0$, for the paired configurations, 
to match the distance and then read
off the free energy.

In the case of the QD configuration, which is the one we will study in
detail, the free energy is given by
$$
F_{QD}=\frac{U_0}{2\pi} \int _1^{\infty} 
\bigg(\frac{\sqrt{z^4-\frac{U_T^4}{U_0^4}}}{\sqrt{z^4-1}} -1\bigg)
-\frac{U_0}{2\pi}+\frac{1}{g}\frac{U_0-2 U_T}{2\pi}
$$
\be
+\sqrt{1+1/g^2}\left(\frac{U_2}{2\pi} \int _{1/\sqrt{x_2}}^{\infty} 
\bigg(\frac{\sqrt{z^4-\frac{U_T^4}{U_2^4}}}{\sqrt{z^4-1}} -1\bigg)
-\frac{U_0}{2\pi}\right)
\ee
and the distance between the quark and the dyon is
$$
L _{QD}= \frac{R^2}{U_0} \sqrt{1-\frac{U_T^4}{U_0^4}} 
\int _1^{\infty} \frac{1}{\sqrt{z^4-\frac{U_T^4}{U_0^4}}\sqrt{z^4-1}}
$$
\be
+\frac{R^2}{U_2} \sqrt{1-\frac{U_T^4}{U_2^4}} 
\int _{1/\sqrt{x_2}}^{\infty} \frac{1}{\sqrt{z^4-
\frac{U_T^4}{U_2^4}}\sqrt{z^4-1}}  .
\ee
We have included the free energy also for the isolated monopole.

\section{Discussion}

We will now compare the free energy for the 
various configurations as the temperature is
increased. 

At zero temperature explicit numerical integration shows that the 
QDM configuration always has
the lowest free energy. 
The physical interpretation is that there is no finite-range
screening. 
The three particles
always feel a Coulomb-like force from the others. Even if one of the
particles is very far from the
other two it is nevertheless advantageous to have a string connecting to
the other particles
rather than
dropping it straight down to the horizon. The last case would have
implied screening. The very long
connecting string does not cost very much since it runs close and
parallell to the horizon which is a
null surface.

It should be clear that the above no-screening result generalizes to 
any configuration.
The exact statement is that, at zero temperature, there can
be {\it only one} type of strings reaching down to $U=0$, all
of the same irreducible $(p,q)$ type ($p,q$ relatively prime).
Indeed, if there were more than one type of stings, we could always 
decrease the action by recombining two strings of different type
some distance $\Delta U$ from $U=0$ into a unique string. (The
gain in the $U$-direction is of order $\Delta U$ while the loss in
the $x_i$-direction is of order $\Delta U^2$.) Reducible strings of
type $(np,nq)$ ($n>0$) can be considered as degenerate cases of $n$ 
overlapping
$(p,q)$ strings. In general, such strings will want to split in order
to globally minimize the action, although we can set up situations where
they do not (e.g., when they all end up on the same $(np,nq)$ particle).
For a singlet configuration, the above means that {\it no} string 
can end on $U=0$. Thus, if there are no singlet
subclusters of particles, all particles are connected to each other 
through a string network and are, therefore, interacting. 

We conclude
that the total electric and magnetic charge cannot be screened at
zero temperature, since the strings from non-singlet clusters
cannot drop to $U=0$ and
will always connect to strings from other non-singlet clusters 
at an arbitrary distance away. The only exception is when all
clusters are of the same $(p,q)$ type, in which case they form a
BPS-saturated state and supersymmetry ensures that there is no net
force between these clusters. Higher multipole moments, on the
other hand, {\it can} in principle be screened since there is nothing
that prevents the strings within a (sufficiently isolated) singlet cluster
to totally close among themselves and prevent the cluster from 
interacting with the rest of the system.

Coming back to the QDM system,
as the temperature is increased, a critical point is eventually reached 
where the free energy for
the QDM configuration coincides with the free energy for
one of the paired configurations.
When the critical point is passed the system is only at a local minimum
of its free energy. 
The system is in a metastable state, and there is a thermal `tunneling'
probability for a transition to the stable state of lower energy.
An estimate of this transition rate can be found by evaluating the
minimal action of a string configuration interpolating between the two
local minima. 

In the particular case described below we have positioned the monopole
some distance away from
the other two. We find that the QD configuration wins at a sufficiently
high temperature. 
The monopole is screened from the rest by the thermal bath. This happens
at free energy less than the one for isolated particles. If the system 
is heated further, the free
energy of the quark-dyon pair approaches the isolated value from below.
When the two are equal, there is a transition to a free system with 
complete screening.

All of these features are illustrated in figure 3 which shows $F_{QDM}$
and
$F_{QD}$ as functions of $U_T$. We have chosen $g=1$
and have subtracted off the free energy of the three isolated particles. 
In the figure $F_{QDM}$ has been drawn for
$y_1 = 0.9$ and $U_0 =1$ while $F_{QD}$ is drawn for $U_0= 0.9$. 
These values correspond to $L_{QD}/R^2$ ranging from $0.9772$ to
$0.8896$ 
for the QDM system and from $0.9655$ to $0.8904$ for the $QD$ system as
$U_T$ goes from $0$ to $0.7$. For the QDM system we have $L_{QM}/L_{QD}$
going from $2.6$ to $2.1$. 
A careful 
readjustment of the parameters to keep $L_{QM}/L_{QD}$ constant would
not change the
conclusions. For the particular configuration that we are considering,
the free energies 
for the QM and DM configurations are always higher and for high enough
temperature 
the configurations do not exist.

\begin{figure}  
\begin{center}                                                         
\epsfig{file=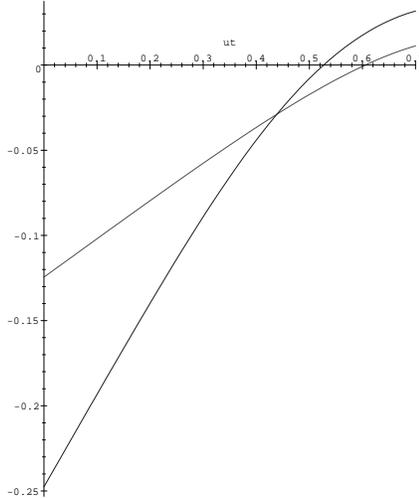,width=7 cm}                                               
     
\caption{ $F_{QDM}$ (steeper curve) and $F_{QD}$ as 
functions of $U_T$.}
\end{center}
\end{figure}

The limit in which some of the particles become much heavier than
the rest is particularly nice. Consider, for instance, the limit
where $g$ becomes very small and thus the monopole and the dyon
acquire a much bigger mass than the quark. Then the AdS solution
for the monopole-dyon system is the same as if the quark did not
exist, since the pull from the quark string is negligible.
This MD solution is essentially the same as for a quark-antiquark
system since the two masses are almost equal and the contribution
of the `leftover' $(-1,0)$ string is negligible. The quark string,
on the other hand, has the option to either go straight to the
horizon or end on the string joining the monopole and the dyon
at an almost right angle at the joint. 
(The preferred configuration will be the one with least action.)
In effect, the light quark
is moving in the background field of the heavy monopole-dyon system.

To make the situation even simpler, consider that the monopole-dyon
distance is bigger than the critical one for screening and therefore
the AdS solution for them corresponds to straight strings down to the
horizon. Their free energy is the same as if they were isolated.
The quark string has the option of either dropping to the
horizon or joining at a right angle either
of these vertical strings (which act, now, essentially like horizons).
In the latter case, the action of the quark string, by symmetry, 
will be half the action of 
a string connecting the quark and an antiquark positioned at the 
mirror image of the quark with respect to the heavy particle. 
So the free
energy of that configuration will be {\it half} the free energy of a
quark-antiquark pair at {\it double} the distance of the quark-heavy
particle pair.

The same reasoning applies when considering a quark-monopole system
at a separation $L$ with the mass of the monopole going to infinity. 
We get for the effective potential $V_{QM} (L,T) 
= F_{QM} (L,T)
- F_Q (T) - F_M (T)$:
\be
V_{QM} (L,T) = \frac{1}{2} V_{Q{\bar Q}} (2L,T)
\ee
at any temperature. This, in particular, explains Minahan's result
that at zero temperature the quark-monopole potential is $1/4$ times
the quark-antiquark potential \cite{Mi}, since at zero temperature the 
potential scales like $1/L$. In the opposite limit of the 
monopole
becoming much lighter than the quark we similarly get $1/4$ times
the monopole-antimonopole potential.

We conclude by mentioning that the above results respect duality
and scale invariance. This can be shown by rescaling $x_i$ and $\tau$
in the metric (1) and appropriately redefining $U$. Specifically, the
free energy $F(Q,L;T,g)$ for {\it any} configuration of a number of static
particles $Q$ at mutual distances $L$ as a function of the temperature $T$
and the coupling constant $g = g_{YM}^2 /4\pi$ obeys
\be
F(Q,L;T,g) = \lambda F(Q, \lambda L; T/\lambda ,g)
= F({\tilde Q},L;T,1/g)
\ee
where $\lambda$ is any (positive) constant and $\tilde Q$ are the
dual particles. In particular, if the particles involved are all
of type $(np,nq)$ for fixed $p,q$ the free energy becomes
\be
F(L,T;p,q,g) = \sqrt{gp^2 + q^2/g} \, F(L;T) 
\ee
which is consistent with duality. (The above is not a BPS-saturated 
state since different $n$ can be both positive and negative.)
One can also verify that in this case the total potential is a sum
of two-body terms.

\end{document}